**Tunneling Spin Injection into Single Layer Graphene**

Wei Han, K. Pi, K. M. McCreary, Yan Li, Jared J. I. Wong, A. G. Swartz, and R. K. Kawakami[†]

Department of Physics and Astronomy, University of California, Riverside, CA 92521

[†] e-mail: roland.kawakami@ucr.edu

**Abstract:**

We achieve tunneling spin injection from Co into single layer graphene (SLG) using $TiO_2$ seeded MgO barriers. A non-local magnetoresistance ($\Delta R_{NL}$) of 130 Ω is observed at room temperature, which is the largest value observed in any material. Investigating $\Delta R_{NL}$ vs. SLG conductivity from the transparent to the tunneling contact regimes demonstrates the contrasting behaviors predicted by the drift-diffusion theory of spin transport. Furthermore, tunnel barriers reduce the contact-induced spin relaxation and are therefore important for future investigations of spin relaxation in graphene.

Spintronics utilizes the electron's spin degree of freedom in addition to its charge in electronic devices for advanced approaches to information storage and processing [1]. Single layer graphene (SLG) is a promising material for spintronics due to the low intrinsic spin-orbit and hyperfine couplings [2], long spin diffusion lengths (~2 μm) [3], and predictions of fascinating spin dependent behavior [4, 5]. Furthermore, SLG is the first material to achieve gate tunable spin transport at room temperature [3, 6, 7]. However, to realize its full potential for spintronics, there are two critical challenges. First, the measured spin lifetimes in SLG (50-200 ps) are orders of magnitude shorter than expected from the intrinsic spin-orbit coupling [2, 3, 8-11]. Consequently, substantial theoretical and experimental effort is focused on identifying the extrinsic mechanism of spin scattering [9, 11, 12]. The second important challenge is to achieve tunneling spin injection into SLG. This will produce efficient spin injection by overcoming the conductance mismatch between the ferromagnetic (FM) metal electrodes and the SLG [13-15]. Up to now, enhancing the spin injection efficiency has focused on reducing the conductance mismatch by decreasing the contact area using MgO masking layers or barriers with pinholes [7, 8, 10, 16-18]. However, tunneling spin injection has not been achieved due to the difficulty to growing uniform, pinhole-free tunnel barriers on graphene.

In this Letter, we demonstrate tunneling spin injection in SLG spin valves and report large spin signals and enhanced spin lifetimes. Using $TiO_2$ seeded MgO films as the tunnel barrier, we observe a non-local magnetoresistance ($\Delta R_{NL}$) as high as 130 Ω at room temperature, which is the largest value observed in any material. The *I-V* characteristics of the contact resistance are highly non-linear and $\Delta R_{NL}$ varies inversely with the SLG conductivity, which are the two principal characteristics of tunneling spin injection. Furthermore, the spin lifetimes (450-500 ps) are considerably longer than previously observed for transparent and pinhole contacts (50-200 ps) [3, 8-10], which suggests that the tunnel barrier greatly reduces the contact-induced spin relaxation. These results are important for applications such as spin-based logic [19] and for fundamental studies of spin relaxation in graphene.

Graphene spin valves are fabricated using mechanically exfoliated SLG flakes on $SiO_2$/Si substrate, where the Si is used as a back gate. Co electrodes are defined by electron-beam lithography using PMMA/MMA bilayer resist to produce undercut, followed by angle evaporation in a molecular beam epitaxy system with a base pressure of $2\times10^{-10}$ torr. Tunneling contacts are fabricated in the following manner. First, 0.12 nm of Ti is deposited at both 0° and 9° angles (Figure 1a), followed by oxidation in $5\times10^{-8}$ torr of $O_2$ for 30 minutes to convert the metallic Ti into insulating $TiO_2$. The presence of $TiO_2$ greatly improves the uniformity of MgO overlayers [20]. A 3 nm MgO masking layer is deposited at an angle of 0° and a 0.8 nm MgO tunnel barrier is deposited at an angle of 9°. Then the 80 nm thick Co electrode is deposited with an angle of 7°. Figure 1b illustrates the geometry of the tunneling contact, where the current flows across the 0.8 nm MgO tunnel barrier of width ~50 nm. Approximately 20% of the tunneling electrodes possess pinholes, which are utilized for investigating the characteristics of spin injection through pinhole contacts. For the transparent contacts, the Co is directly contacted to SLG with a 2 nm MgO masking layer [7, 8].

Spin injection and transport are measured on samples held at 300 K in helium atmosphere using the non-local geometry with standard ac lock-in techniques [3, 21]. The inset of Figure 1c shows the non-local measurement where the spin is injected at electrode E2 and detected at E3. The non-local resistance, $R_{NL}$, is defined as the measured voltage signal ($V_{NL}$) divided by the injection current (*I*). Figure 1c shows $R_{NL}$ as the magnetic field is swept up (black curve) and swept down (red curve) for a device with tunneling contacts. $\Delta R_{NL}$ is defined as the difference of $R_{NL}$ between the parallel and antiparallel magnetization states of E2 and E3. For spin transport across the 2.1 μm electrode gap ($L$), $\Delta R_{NL}$ is 130 Ω (Figure 1c), which is the largest value observed in any lateral spin valve including metals and semiconductors [3, 6] [7, 16, 22]. For tunneling contacts [3]

$$\Delta R_{NL} = \frac{1}{\sigma_G} \frac{P_J^2 \lambda_G}{W} e^{-L/\lambda_G} \qquad (1)$$

where $P_J$ is the spin injection/detection efficiency, and $\sigma_G$, $W$, and $\lambda_G$ are the conductivity, width, and spin diffusion length of the SLG, respectively. $P_J$ is calculated to be 26-30% using experimental values of $\sigma_G$ = 0.35 mS, $W$ = 2.2 μm, $L$ = 2.1 μm, and typical values of $\lambda_G$ = 2.5-3.0 μm (see Figure 4). This compares favorably with the tunneling spin polarization of 35% measured by spin-dependent tunneling from Co into a superconductor across polycrystalline $Al_2O_3$ barriers [23]. The spin injection efficiency is larger than observed in previous studies using barriers with pinholes (2% – 18% at low bias) [3, 10, 16] and transparent contacts (1%) [8].

The tunnel barrier enhances the efficiency of spin injection from Co into the SLG by alleviating the conductance mismatch problem [13-15]. For spin injection without tunnel barriers, the spins that are injected from the Co electrode into the SLG can diffuse within the SLG (toward neighboring electrodes) or diffuse back into the Co electrode. The flow of spin via diffusion is governed by the spin resistances [18] which are $R_G = \lambda_G/(\sigma_G W)$ for the SLG, and $R_F = \rho_F \lambda_F / A_J$ for the Co, where $\rho_F$ is the Co resistivity, $\lambda_F$ is the spin diffusion length of Co, and $A_J$ is the junction area [24]. Using typical parameters ($W$ = 2 μm, $\lambda_G$ = 2-3 μm, $\sigma_G$ = 0.5 mS, $\rho_F$ = $6 \times 10^{-8}$ Ω m [25], $\lambda_F$ = 0.06 μm [26]), the $R_F/R_G$ ratio has values between ~$10^{-3}$ and ~$10^{-5}$ depending on the value of $A_J$ [24]. Because $R_F \ll R_G$, the spin diffusion is dominated by the back flow of spins into the Co electrode, which leads to a low spin injection efficiency. The insertion of a tunnel barrier increases the spin injection efficiency by blocking the back flow of spins into the Co electrode.

Quantitatively, the role of the tunnel barrier is explained in the one-dimensional drift-diffusion theory of spin transport [18], where $\Delta R_{NL}$ is given by

$$\Delta R_{NL} = 4R_G e^{-L/\lambda_G} \left( \frac{P_J \frac{R_J}{R_G}}{1-P_J^2} + \frac{P_F \frac{R_F}{R_G}}{1-P_F^2} \right)^2 \times \left( \left(1 + \frac{2\frac{R_J}{R_G}}{1-P_J^2} + \frac{2\frac{R_F}{R_G}}{1-P_F^2}\right)^2 - e^{-2L/\lambda_G} \right)^{-1} \quad (2)$$

where $P_F$ is the spin polarization of the FM and $R_J$ is the contact resistance between the FM and SLG. This equation shows that increasing the contact resistance produces a strong enhancement of $\Delta R_{NL}$ that saturates as $R_J$ becomes significantly larger than $R_G$ [see supplemental information]. In addition to the magnitude of $\Delta R_{NL}$, another method to distinguish the tunneling contacts is to investigate the relationship between $\Delta R_{NL}$ and $\sigma_G$, which can be tuned by gate voltage. For tunneling contacts $\Delta R_{NL}$ scales with $1/\sigma_G$ (equation 1), while for transparent contacts ($R_J << R_G$) $\Delta R_{NL}$ scales with $\sigma_G$ [7, 11]. Figure 2 shows the calculated gate dependence of $\Delta R_{NL}$ for transparent, intermediate ($R_J \sim R_G$), and tunneling contacts [see supplemental information]. For transparent contacts, the linear increase of $\Delta R_{NL}$ with gate voltage is due to the linear increase of $\sigma_G$ away from the Dirac point [27]. For tunneling contacts, $\Delta R_{NL}$ varies inversely with gate voltage and exhibits a peak at the Dirac point.

Figures 3a and 3b show the experimental results for the gate dependence of $\Delta R_{NL}$ for SLG spin valves with transparent and pinhole contacts, respectively. The *I-V* characteristic of the contact resistance is determined by a three-probe lock-in measurement (current is applied across E1 and E2, voltage measured across E3 and E2). For both cases, the nearly constant bias dependence of $(dV/dI)_C$ (insets of Figures 3a and 3b) corresponds to a nearly linear *I-V* characteristic. For transparent contacts, $\Delta R_{NL}$ (black squares) exhibits a minimum at the Dirac point, and a linear relationship with $\sigma_G$ (red curve), which verifies the theoretical prediction (Figure 2, top curve). For pinhole contacts, $\Delta R_{NL}$ (black squares) shows relatively little variation and has a weak minimum near the Dirac point which is similar to the case of intermediate contact resistance as calculated in Figure 2 (middle curve). For both the transparent and pinhole contacts, the non-local MR and *I-V* characteristics are consistent with previous studies [3, 7, 8, 10, 16] which exhibit a minimum in $\Delta R_{NL}$ at the Dirac point and nearly linear I-V curves for the contacts.

For tunneling contacts, the *(dI/dV)$_C$* is highly non-linear (Figure 3c inset) and exhibits little temperature dependence, which are consistent with tunneling transport across the Co/MgO/TiO$_2$/SLG junctions. Figure 3c shows the gate dependence of $\Delta R_{NL}$ (black squares) and $\sigma_G$ (red curve) for tunneling contacts. Interestingly, $\Delta R_{NL}$ exhibits a maximum at $V_g$ = 2 V near the Dirac point, which is the first time this has been observed experimentally. The origin of the asymmetry of $\Delta R_{NL}$ vs. $V_g$ is unclear and varies from sample to sample. The observed peak structure in the gate dependence is a key characteristic of tunneling spin injection (Figure 2, bottom curve), and has been reproduced on four different devices. This inverse scaling of $\Delta R_{NL}$ with $\sigma_G$ is associated with the spin injection process as opposed to spin detection. Specifically, spin injection produces a difference in the spin-dependent chemical potential at the tunnel barrier/SLG interface given by $\Delta \mu = \mu_\uparrow - \mu_\downarrow = eP_J R_G I$ [18]. Thus, a larger $R_G$ will increase $\Delta R_{NL}$ due to a greater difference in the spin-dependent chemical potential.

While the achievement of tunneling spin injection will be important for applications in spintronics, it will also have a strong impact on fundamental studies of spin relaxation in graphene. As shown in Figures 4a and 4b, the spin lifetimes measured at the Dirac point are 495 ps and 448 ps for tunneling SLG spin valves with 2.1 μm and 5.5 μm spacing, respectively. These are much longer than the spin lifetimes of 134 ps for pinhole contacts (Figure 4c) and 84 ps for transparent contacts (Figure 4d), which are consistent with the values reported in previous studies (50 - 200 ps) [3, 8-10]. The spin lifetimes are obtained by applying an out-of-plane magnetic field ($H_\perp$) to induce spin precession and fitting the resulting Hanle curves (see [11] for details) with

$$R_{NL} \propto \pm \int_0^\infty \frac{1}{\sqrt{4\pi Dt}} \exp\left[-\frac{L^2}{4Dt}\right] \cos(\omega_L t) \exp(-t/\tau_s) dt \qquad (3)$$

where the + (-) sign is for the parallel (antiparallel) magnetization state, $D$ is the diffusion constant, $\tau_s$ is the spin lifetime, and $\omega_L = g\mu_B H_\perp / \hbar$ is the Larmor frequency. Theoretically, the measured spin lifetime ($\tau_s$) is determined by the spin-flip scattering within the SLG (at a rate of $\tau_{sf}^{-1}$) and spin relaxation induced by the Co contacts. In the latter effect, the spins diffuse into the Co contact with characteristic escape time ($\tau_{esc}$), which limits the measured spin lifetime. For $\lambda_G \rightarrow \infty$, these time scales are simply related by $\tau_s^{-1} = \tau_{sf}^{-1} + \tau_{esc}^{-1}$ [28], while for the more realistic case of finite $\lambda_G$, the influence of the contact-induced relaxation should be reduced. Furthermore, spin-flip scattering at the Co/SLG interface may introduce additional spin relaxation. Due to the increased spin lifetimes, the spin diffusion lengths from the Hanle fits ($\lambda_G = \sqrt{D\tau_s}$) are significantly larger for tunneling contacts (2.5-3.0 μm) than for transparent and pinhole contacts (1.2-1.4 μm). The longer spin lifetimes and spin diffusion lengths with tunneling contacts indicate that the effect of the contact-induced relaxation is substantial for transparent and pinhole contacts. Thus, tunnel barriers reduce the contact-induced relaxation and enable a more accurate measurement of $\tau_{sf}$ for fundamental studies of spin relaxation.

In conclusion, we have successfully achieved tunneling spin injection into SLG using $TiO_2$

seeded MgO barriers and observe enhanced spin injection efficiencies and large $\Delta R_{NL}$. Investigating $\Delta R_{NL}$ vs. $\sigma_G$ for the different contact regimes (from transparent to tunneling) realizes the contrasting behaviors predicted by the drift-diffusion theory. Finally, tunnel barriers reduce the contact-induced spin relaxation and are therefore important for future investigations of spin relaxation in graphene.

We acknowledge stimulating discussions with E. Johnston-Halperin and acknowledge the support of ONR (N00014-09-1-0117), NSF (CAREER DMR-0450037), and NSF (MRSEC DMR-0820414).

FIGURE CAPTIONS:

Figure 1: (a) Schematic diagram of the angle evaporation geometry. The grey layers are PMMA/MMA resist with undercut. The red and blue dashed lines show the 0° and 9° deposition of $TiO_2$ and MgO. The black lines indicate the 7° evaporation of Co. (b) Schematic drawing of the Co/MgO/$TiO_2$/SLG tunneling contacts. The arrow indicates the current flow through the MgO tunnel barrier. (c) Non-local MR scans of a SLG spin valves measured at room temperature. The black (red) curve shows the non-local resistance as the magnetic field is swept up (down). The non-local MR ($\Delta R_{NL}$) of 130 Ω is indicated by the arrow. Inset: the non-local spin transport measurement on this device with a spacing of $L$ = 2.1 μm and SLG width of $W$ = 2.2 μm.

Figure 2: Predictions of the drift-diffusion theory of spin transport. The non-local MR as a function of gate voltage for three different types of contacts between Co and SLG: transparent, intermediate, and tunneling. The curves are normalized by their value at zero gate voltage.

Figure 3: (a-c) Non-local MR (black squares) and conductivity (red lines) as a function of gate voltage for SLG spin valves with transparent, pinhole and tunneling contacts, respectively. Inset: the differential resistance of the contact, $(dV/dI)_C$, as a function of bias current.

Figure 4: (a) Hanle spin precession for SLG spin valves with tunneling contacts ($R_J$ = 30-70 kΩ, non-linear) for $L$=2.1 μm. (b) Hanle spin precession for tunneling contacts ($R_J$ = 20-40 kΩ, non-linear) with $L$=5.5 μm. (c) Hanle spin precession for pinhole contacts ($R_J$ = 6 kΩ, linear) with $L$=2.0 μm. (d) Hanle spin precession for transparent contacts ($R_J$ < 0.3 kΩ, linear) with $L$=3.0 μm. The top (red/grey) and bottom (black) curves correspond to Hanle curves of the parallel and anti-parallel states, respectively. The solid lines are best fit curves based on equation 3. The units for $D$ are $m^2/s$.

Figure 1

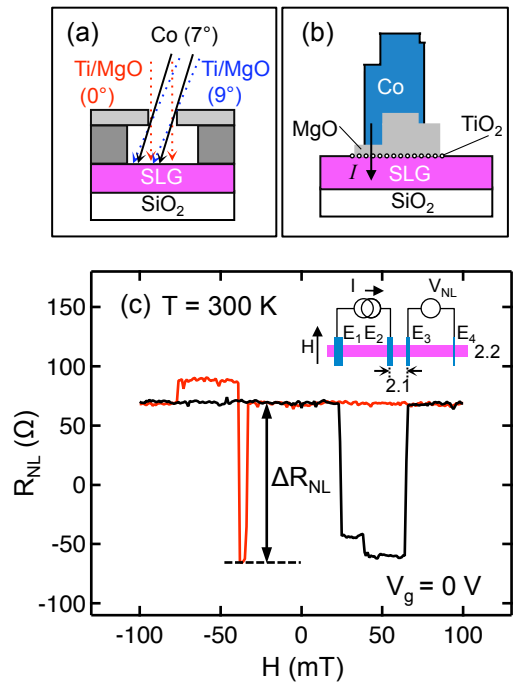

Figure 2

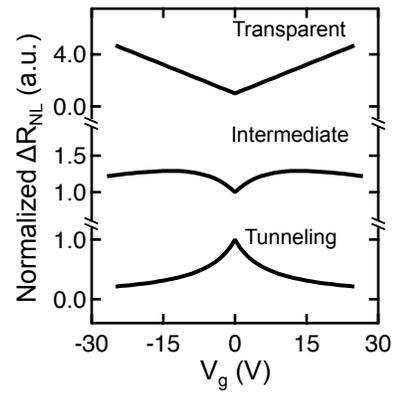

Figure 3

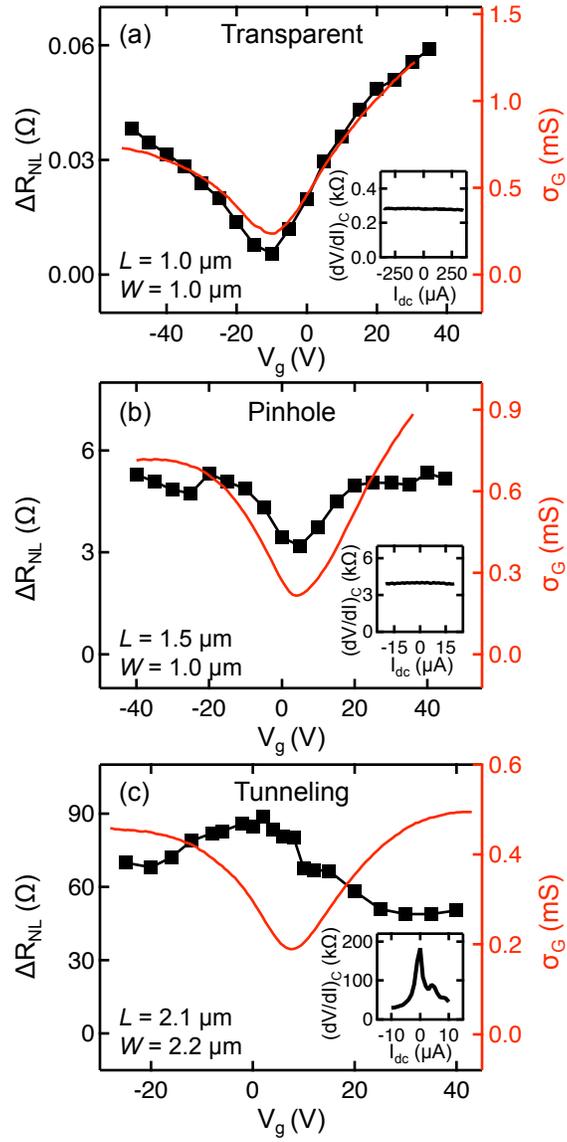

Figure 4

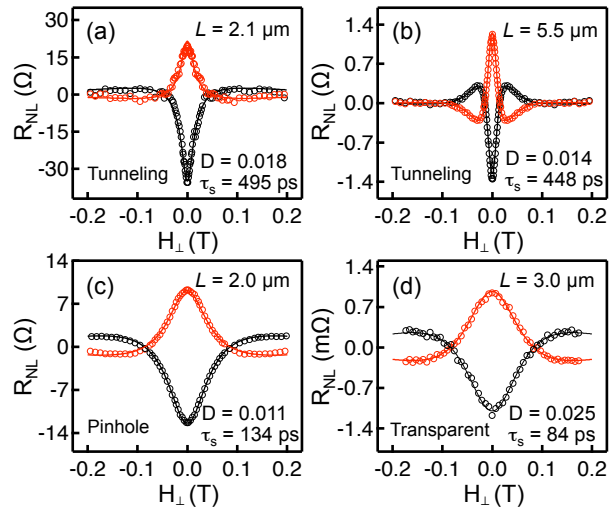